\documentclass[twocolumn, times, twocolappendix]{aastex631}
\usepackage{amsmath}
\usepackage{graphicx,ulem}
\usepackage{xcolor}
\newcommand{\gf}{\bf \color{blue}}
\def\SDO{{\textit{SDO}}}
\def\mw{{microwave}}

\def\gr{{gyroresonant}}

\received{\today}
\revised{*}
\accepted{*}
\submitjournal{ApJ}

\shorttitle{Coronal heating law}
\shortauthors{Fleishman et al.}

\begin{document}
 \title{
Steady-State Heating of Diffuse Coronal Plasma in a Solar Active Region
 }

\correspondingauthor{Gregory D. Fleishman}
\email{gfleishm@njit.edu}

\author[0000-0001-5557-2100]{Gregory D. Fleishman}
	\affil{Center For Solar-Terrestrial Research, New Jersey Institute of Technology, Newark, NJ 07102, USA}
    \affil{Institut f\"ur Sonnenphysik (KIS), Georges-Köhler-Allee 401 A, D-79110 Freiburg, Germany}

\author[0000-0001-8644-8372]{Alexey A. Kuznetsov}
\affiliation{Institute of Solar-Terrestrial Physics, Irkutsk 664033, Russia}

\author[0000-0003-2846-2453]{Gelu M. Nita}
	\affil{Center For Solar-Terrestrial Research, New Jersey Institute of Technology, Newark, NJ 07102, USA}    



\begin{abstract}
Solar corona is much hotter than lower layers of the solar atmosphere---photosphere and chromosphere. The coronal temperature is up to 1\,MK in quiet sun areas, while up to several MK in active regions, which implies a key role of  magnetic field in coronal heating. This means that understanding coronal heating requires reliable modeling of the underlying three-dimensional (3D) magnetic structure of an active region validated by observations. Here we employ synergy between 3D modeling, optically thick gyroresonant \mw\ emission, and optically thin EUV emission to (i) obtain and validate the best magneto-thermal model of the active region and (ii) disentangle various components of the EUV emission known as diffuse component, bright loops, open field regions, and ``moss'' component produced at the transition region. Surprisingly, the best thermal model corresponds to high-frequency energy release episodes, similar to a steady-state heating. Our analysis did not reveal significant deviations of the elemental abundances from the standard coronal values.
\end{abstract}

\section{Introduction}

Why solar corona is much hotter than the underlying photosphere and chromosphere and what specific physical mechanisms are responsible for that are intriguing fundamental unanswered questions in solar physics. The corona is hot in both quiet sun areas and above active regions (ARs), while the AR plasma is roughly three times hotter than the quiet sun one. This implies that the magnetic field, which has more energy in ARs than in quiet Sun areas, plays a key role in coronal heating. However, traditionally, most of the information on the coronal thermal plasma has been obtained from extreme ultraviolet (EUV) or soft X-ray (SXR) emissions, which are not explicitly sensitive to the magnetic field. 

\begin{figure*}[!th]
\centerline{%
\includegraphics[width=0.999\linewidth]{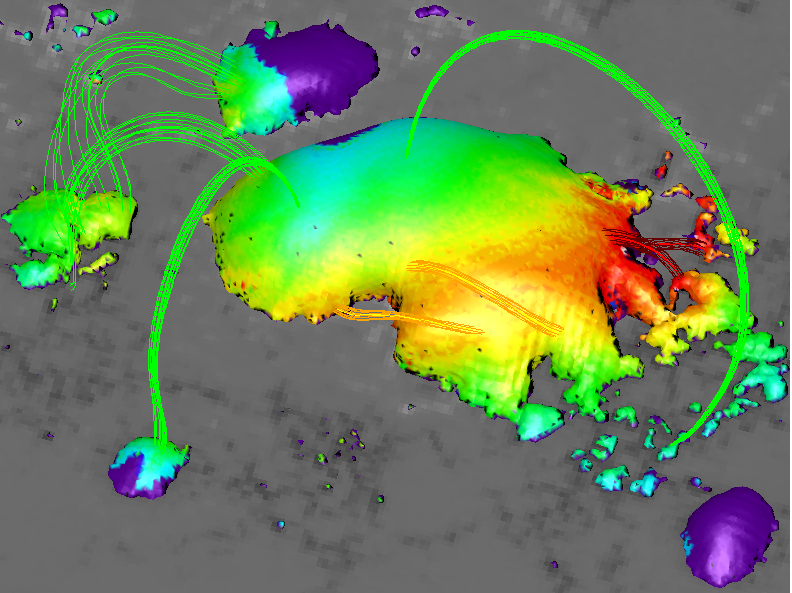}}%
\caption{Illustration of complementarity of the \mw\ and EUV emission. The colored surface is the isogauss surface of the magnetic field $B=678$\,G, obtained from our NLFFF model of AR 11520, that corresponds to the third harmonic of the \gr\ emission at 5.7\,GHz. The color shows distribution of the plasma temperature over the surface. Several color-coded flux tubes are shown to illustrate one-dimensional character of the heat transfer in the magnetic-field-dominated corona: green lines cross the surface at green areas at both ends, yellow in yellow, and red in red. One can see that the temperature does not change much along the field lines, while it changes noticeably along the isogauss surface (accross the field lines). This means that reproducing a correct distribution of the plasma temperature along an isogauss surface requires a correct heating model over a range of magnetic flux tubes with a range of lengths and magnetic field strengths.
}
\label{Fig:GR_layer_lines}
\end{figure*}

{
A complementary, magnetic-field-sensitive diagnostic of the coronal plasma is available from the \mw\ emission of the thermal plasma in the ambient magnetic field, called the gyroresonant (GR) emission. Figure\,\ref{Fig:GR_layer_lines} illustrates this complementarity by showing a combination of magnetic field lines / flux tubes and a 3D surface of equal magnetic field value (678 G isogauss surface). Indeed, 
the heat transfer occurs along the magnetic flux tubes because the heat conduction is highly anisotropic 
in the magnetic-dominated corona. Thus, the magnetic flux tubes are roughly isothermal which results  in loop-like EUV or SXR features. In contrast, the \gr\ emission is optically thick at narrow gyro layers inscribing the isogauss surfaces and, as so, at a given frequency, the \mw\ emission mainly comes from a surface of equal magnetic field (isogauss surface), whose temperature, however, is not constant, but varies according to the plasma temperature in the magnetic flux tubes crossing the gyrolayer. This means that the \mw\ emission from a given gyrolayer samples different temperature regimes according to the magnetic field lines crossing it; this is the fundamental reason for the \mw\ emission sensitivity to both magnetic and thermal AR structure.
}

Recently, \citet{Fleishman2021a} have analyzed the coronal heating processes using microwave observations of the Siberian Solar Radio Telescope (SSRT) and the Nobeyama Radioheliograph (NoRH) at the frequencies of 5.7 and 17 GHz, respectively, in  AR 11520 on 2012-07-12. 
\citet{Fleishman2021a} discovered that the synthetic images generated from a 3D magnetic model ``dressed'' with a parametric thermal model \citep{Nita2018,Nita2023} are highly sensitive to specific parameters of this thermal model and, thus, this sensitivity can be used to identify the best heating regime, or a set of such regimes. 

However, the study of \citet{Fleishman2021a} had several important limitations: (i) they employed a simplified treatment of the gyroresonance emission using temperature and density values estimated as moments of the modeled DEM distributions, which introduces an unknown bias; (ii) they explored only a very limited parameter space; (iii) they explored only one (impulsive) nanoflare heating model; and (iv) considered only one instance of one AR. In this work, we remedy this situation by (i) using a new theory and associated computer codes \citep{Fleishman2021b} that explicitly take into account the multi-temperature plasma in each elementary model element instead of using the distribution moments; (ii) employing systematic searches over an extended parameter space; (iii) exploring all currently available coronal heating lookup tables \citep{Nita2023}, while deferring an analysis of other ARs to a forthcoming dedicated study. 

\section{Observations}
The instruments and data employed for the study of AR 11520 are described in \citet{Fleishman2021a}. Namely, we use microwave intensity images obtained around local noons by the SSRT \citep{Grechnev2003} and NoRH \citep{Nakajima1994} at the frequencies of 5.7 and 17 GHz, respectively. We also use contextual \SDO/AIA \citep{Lemen2012} EUV images at different wavelengths and the \SDO/HMI \citep{Scherrer2012} data, the later being used to reconstruct the magnetic field structure in the corona.

\section{Parametric Coronal Heating }
\label{S_PCHM}

The three-dimensional (3D) AR modeling framework is extensively described by \citet{Nita2018,Nita2023}. Here, we briefly outline the employed methodology, referring to \citet{Nita2023} for details. The 3D modeling is based on the magnetic data cube obtained from nonlinear force-free field (NLFFF) extrapolation of the photospheric vector magnetic field. This data cube is dressed with a chromospheric plasma model and a {tunable} parametric coronal heating model{\gf, described below}. These steps are packaged together within the Automated Model Production Pipeline \citep[AMPP,][]{Nita2023}.

{As we have explained in the Introduction, in the magnetic-field-dominated corona, the heat is transferred predominantly along the magnetic field; thus, simplified one-dimensional hydrodynamic models can be applied to individual field lines with finite lengths (closed ones). Theoretical ideas about the coronal heating \citep[e.g.,][]{2015RSPTA.37340256K} suggest that the volumetric coronal heating rate $Q$ (erg $\textrm{cm}^{-3}$ $\textrm{s}^{-1}$) may depend on parameters of the corresponding magnetic flux tube such as the magnetic field strength, coronal loop length, mass density, photospheric velocity, loop radius, curvature, or twist, but the fundamental ones are the first two, with the others parametrizable in terms of them.}
The{refore}, for the volume elements associated with closed field lines, we define  the volumetric heating rate in the following parametric form
\begin{equation}\label{HeatingRate}
Q=Q_0\left(\frac{\left<B\right>}{B_0}\right)^a\left(\frac{L}{L_0}\right)^{-b},
\end{equation}
where $\left<B\right>$ is the magnetic field strength averaged along the field line, $L$ is the length of the field line, $B_0$ and $L_0$ are  normalization constants chosen to be 100 G and $10^9$ cm, respectively, $Q_0$ characterizes the heating intensity, and the power-law indices $a$ and $b$ depend on the physical heating mechanism \citep[see, e.g.,][]{Mandrini2000}. {If $Q_0$ is the same for all voxels, the thermal model will produce some mean properties of the diffuse coronal component, while reproducing individual bright EUV loops may require proportionally enhanced $Q_0$ in a subset of the coronal structures.}
To apply the parametric heating model to the magnetic skeleton, 
AMPP computes the magnetic field lines passing through the center of each volume element in the corona and saves $\left<B\right>$ and $L$ as voxel's properties. 

To associate a given volumetric heating rate with the thermal plasma properties of a given voxel, we employ hydrodynamical simulations, using the Enthalpy-Based Thermal Evolution of Loops code \citep[EBTEL,][]{Klimchuk2008, Cargill2012a, Cargill2012b, Barnes2016}. This code simulates plasma heating/cooling processes, accounting for chromospheric evaporation and condensation. For a given heating rate $Q$ and magnetic flux tube length $L$, the EBTEL code computes the corresponding equilibrium characteristics of thermal plasma, the differential emission measure (DEM) and differential density metric \citep[DDM,][]{Fleishman2021b}, {in both the coronal volume and the transition region (TR), the latter treated as a narrow surface at the bottom of the corona}. The GX Simulator package currently contains eight precomputed lookup tables of the EBTEL code outputs corresponding to different heating regimes \citep{Nita2023}. {They include (a) a steady-state lookup table (computed as an outcome of many frequent low-amplitude heating events), (b) a generic impulsive one that assumes a fixed frequency of one nanoflare per 10,000\,s, and (c) six lookup tables with ``adaptive'' high ($\tau=0.2\tau_0$), moderate ($\tau=1\tau_0$), and low ($\tau=5\tau_0$) frequencies of nanoflares, $1/\tau$, scaled with the cooling time $\tau_0$ of the given loop, and power-law distributions of nanoflare  amplitudes with indices $-1$ (dominated by large events) and $-2.5$ (dominated by small events).}

{The DEM distributions are used to compute EUV, SXR, and free-free radio emissions from the coronal and TR voxels. The DDM distributions are needed to compute GR radio emission from the coronal volume, while the TR does not contribute significantly to this component. }
As demonstrated by \citet{Fleishman2021b}, GR emission from a plasma composed of multiple thermal components may differ considerably from the emission from a single-temperature plasma with the same average density and temperature (the emission from a multi-temperature plasma is noticeably brighter), and hence considering the DDM distributions is very important for an accurate emission synthesis.

The {available EBTEL lookup tables}, together with the 3D magnetic field structure, are used to {synthesize the AR thermal emissions at various heating regimes. Then,}
the model images are  convolved with the instrumental point-spread functions (PSFs) to enable meaningful quantitative comparison with observations {and, based on this comparison, the model parameters are adjusted to achieve the best possible agreement between the model and observations. 
Together, the above steps are packaged together in a
coronal heating modeling pipeline (CHMP).}

\begin{figure*}
\centerline{%
\includegraphics[width=0.999\textwidth]{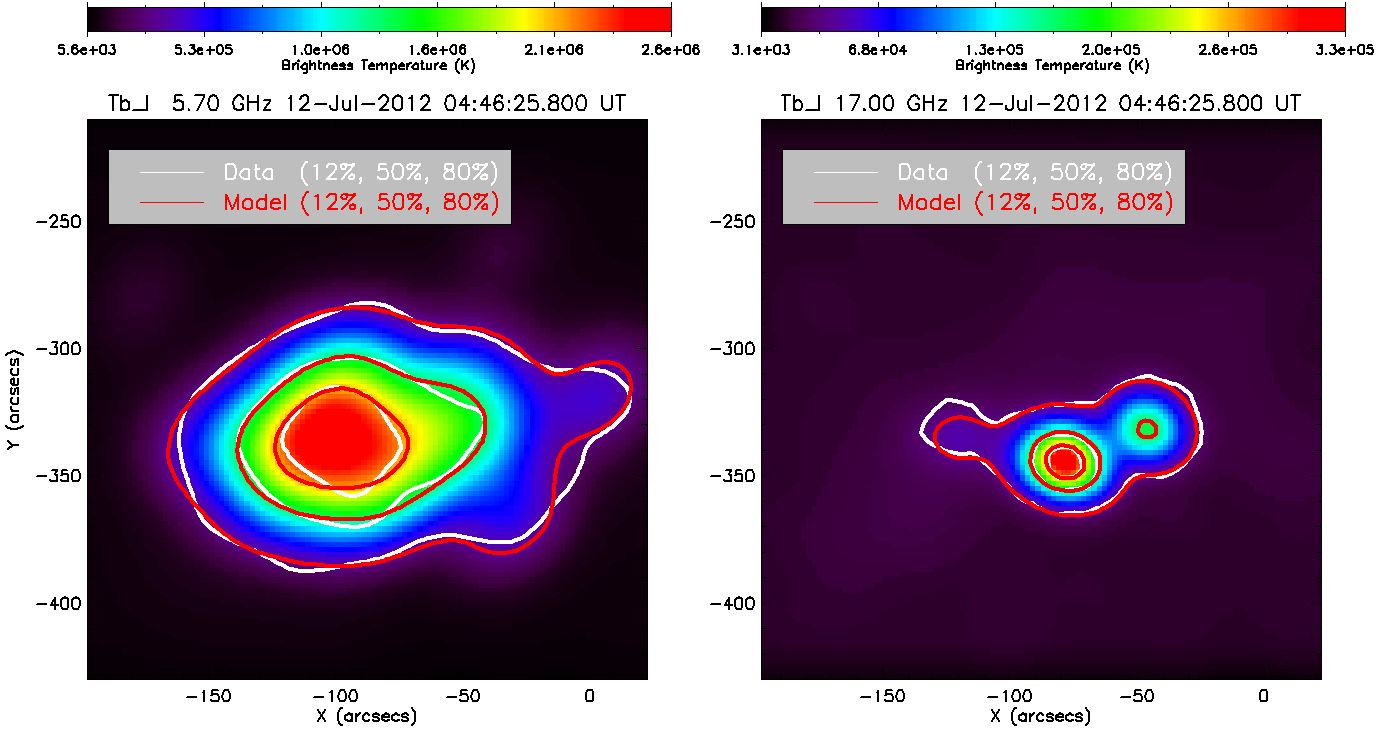}}%
\caption{Comparison of synthetic and observed microwave images of AR 11520 at 5.7 and 17 GHz.  
Synthetic maps are shown as the back-ground images and red contours at 12, 50, and 80\%. Observations are shown as white contours at the same levels. The 12\% level was selected as a threshold for the model fine-tuning. The synthetic maps are produced for the best thermal model obtained with the CHMP for the time of the model creation shown in the panel titles; the observational maps were rotated to this time frame. The model and data show remarkable morphological similarity quantified with cross-correlation coefficients of about 99.5\%.}
\label{Fig:images}
\end{figure*}

\begin{figure*}[!th]
\centerline{%
\includegraphics[width=0.999\textwidth]{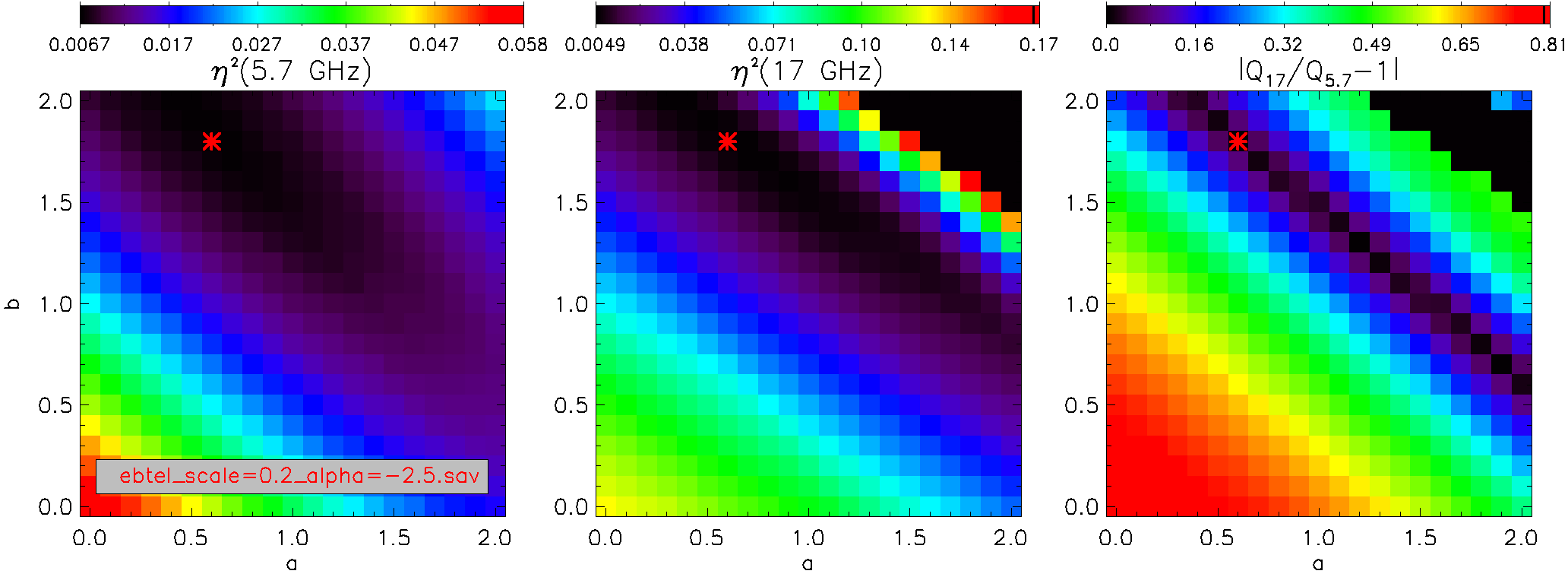}}\centerline{\includegraphics[width=0.999\textwidth]{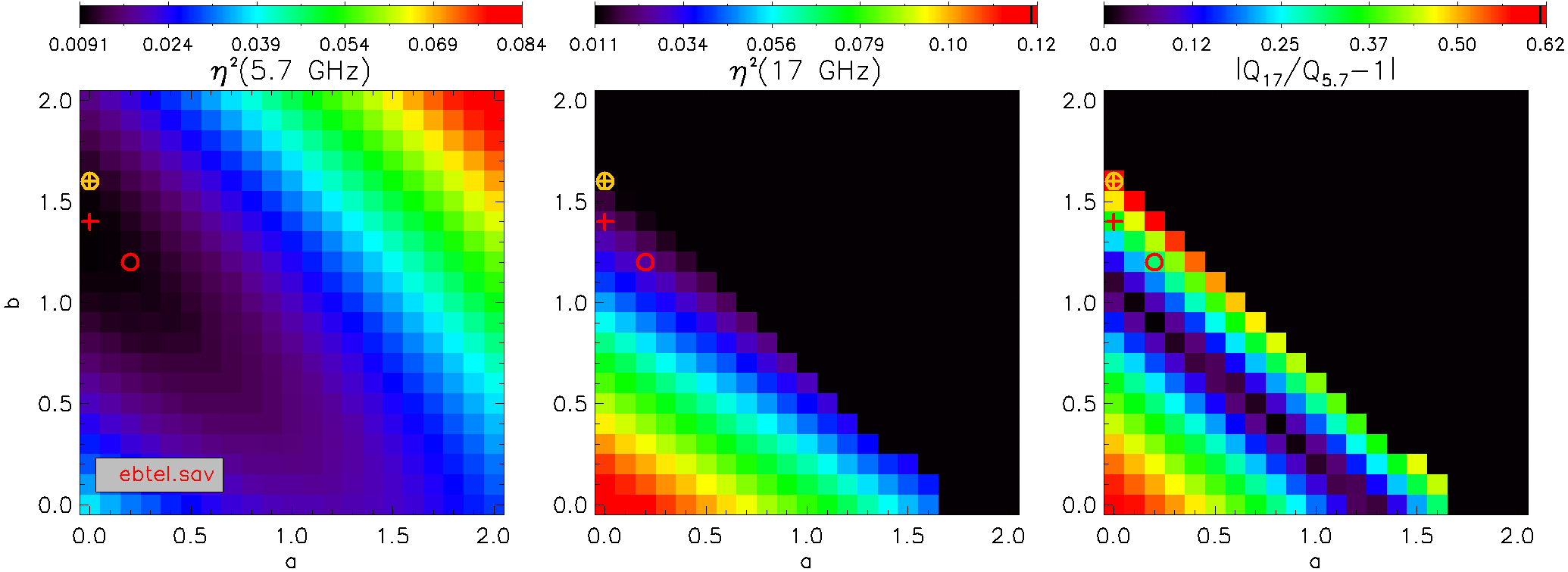}}%
\caption{Selection of the best magneto-thermal model of AR 11520 is shown in the top row, where the left and middle panels show the maps of the normalized residual metrics $\eta^2$ at 5.7 and 17 GHz, respectively, for the search over the \texttt{‘ebtel\_scale=0.2\_alpha=-2.5.sav’} EBTEL lookup table. This table was  found to offer a better match to the data compared with all other lookup tables. Red asterisks point at the best $(a, b)$ combination---the same at both frequencies and both metrics, i.e., $\eta^2$ and cross-correlation coefficient $CC$. The right panel shows a mismatch between the heating normalization constants $Q_0$ needed to reproduce the emission at two considered frequencies; note a non-zero mismatch in a general case of arbitrary ($a, b$) pair, while a valid solution must have no mismatch in $Q_0$. The selected solution satisfies this natural requirement. The bottom row shows similar search results for the impulsive heating lookup table \texttt{`ebtel.sav'}, where the best solutions at different frequencies do not coincide with each other; the best $\eta^2$ metrics are shown in crosses, while the best $CC$---by the empty circles, in red for 5.7 and in yellow for 17\,GHz. All these solutions are off the line of equal $Q_0$ for both frequencies as apparent from the right panel. The black areas in the middle panels are where the CHMP failed to find a solution for 17\,GHz; these areas are anyway not favored by the 5.7\,GHz data.}
\label{Fig:metrics}
\end{figure*}

\section{Coronal Heating Modeling Pipeline}
\label{S_CHMP}

To perform an automated search of the best coronal heating model, we employ the CHMP introduced by \citet{Nita2023}, which has been implemented as a part of the GX Simulator package and as a standalone code\footnote{\url{https://github.com/kuznetsov-radio/gxmodelfitting}}; {version 1.0.0 is archived in Zenodo \citep{Kuznetsov2025}}. The pipeline permits minimization of a chosen goodness-of-the-fit parameter; for this study we selected the normalized residual $\eta^2$ defined as 
\begin{equation}\label{EtaMetric}
\eta^2=\frac{\left<(I_{\mathrm{obs}}-I_{\mathrm{mod}})^2\right>}{\left<I_{\mathrm{obs}}\right>^2},
\end{equation}
where $I_{\mathrm{obs}}$ is the observed brightness and $I_{\mathrm{mod}}$ is the synthetic (model) brightness; all these values are attributed to individual pixels and are dependent on the image coordinates $x$ and $y$. All averagings in Equation (\ref{EtaMetric}) are performed over the area where either observed or synthetic microwave brightnesses are sufficiently high to satisfy the condition 
\begin{displaymath}
I_{\mathrm{obs}}>\alpha\max(I_{\mathrm{obs}})\quad\textrm{or}\quad I_{\mathrm{mod}}>\alpha\max(I_{\mathrm{mod}});
\end{displaymath} 
we used the threshold value of $\alpha=0.12$, the same as in \citet{Fleishman2021a}.

In this study, we performed a systematic search for the best solution {for the diffuse thermal plasma component of the coronal volume (that is, excluding TR)} by varying the free parameters $Q_0$, $a$ and $b$ within the eight available EBTEL lookup tables. {These best solutions are aimed to reproduce the \mw\ images which are sensitive to the coronal plasma but insensitive to the TR}. For each lookup table, we iterate over a preselected grid of $a$ and $b$ values to identify the best $Q_0$ that provides the best (lowest) $\eta^2$ metrics for the given $(a, b)$ pair at each frequency, i.e., at 5.7 GHz for the SSRT observations and 17 GHz for the NoRP observations; we also compute the corresponding cross-correlation coefficients $CC$ between the observed and synthetic images. Then, we iterate over all EBTEL lookup tables, and finally we select the solution, where the best $\eta^2$ and $CC$ metrics are achieved at both considered frequencies for the same $Q_0$, $a$, and $b$ combination. The EUV emission from the obtained best-fit model are then computed using the GX Simulator tool \citep{Nita2023}, with various contributions of the transition region (see Section \ref{AIAvalidation}).

\section{Results}



To determine the best parametric heating model, we systematically inspected a grid of indices $0\le a\le 2$, $0\le b\le 2$, which we proved sufficient to identify the best solutions for all of the available EBTEL lookup tables for the analyzed instance of AR 11520. 
The comparison between the best synthetic and observed images for the model described below is shown in Figure\,\ref{Fig:images}. A visual inspection of the synthetic emission (red contours) reveals extremely good agreement with the data (white contours), quantified by cross-correlation coefficients of about 99.5\% at both frequencies. 
Typically, for a given frequency and given lookup table the best solutions occupy an elongated area of $(a, b)$ pairs indicating certain ``degeneracy lanes'' of appropriate solutions, see Figure\,\ref{Fig:metrics}. 

In addition, we compared the best values of the typical volumetric heating rates needed to reproduce emission at two different frequencies and found that they are generally different from each other, while a valid model must reproduce emission at all frequencies with the same volumetric heating rate. We discovered that for more impulsive (with low-frequency nanoflares) heating models, the best solutions at different frequencies do not agree, and that the residual metrics are worse compared with models implying a more steady (high-frequency) heating. Comparing the search results, as illustrated in Figure\,\ref{Fig:metrics}, we found that the data are best reproduced by the lookup table \texttt{‘ebtel\_scale=0.2\_alpha=-2.5.sav’} 
\citep{Nita2023}, i.e., by a high-frequency heating model with the median time between heating events (nanoflares) $\tau=0.2 \tau_0$ 
dominated by small events (the heating event amplitude distribution index is $-2.5$). The best model is marked with red asterisks in the top row of Figure\,\ref{Fig:metrics}, which points to a parameter combination $a=0.6$, $b=1.8$ that provides both the minimum residuals, maximum cross-correlation coefficients, and the same heating rate constant $Q_0$ at both frequencies. The two frequencies show two slightly different degeneracy ``lanes'': $b=2.4-a$ at 5.7\,GHz and $b=2.2-0.7a$ at 17\,GHz which intersect at our best solution $a=0.6$ and $b=1.8$.

The second best model is the legacy steady-state EBTEL lookup table \texttt{`ebtel\_ss.sav'} \citep{Nita2023}. The overall picture of the $a-b$ diagrams (not shown) is similar to that shown in Figure\,\ref{Fig:metrics}, top row, while the details are different. The two frequencies show two distinct degeneracy lanes: $b=3.04-1.33a$ at 5.7\,GHz and $b=2.3-0.75a$ at 17\,GHz which intersect at roughly $a=1.28$ and $b=1.34$. {The best-fit $(a, b)$ pairs are of about $(0.8, 1.8)$ at 5.7 GHz and $(0.6, 1.8)$ at 17 GHz, which are very close to each other, but deviate slightly from the intersection point of the degeneracy lanes.} The cross-correlation coefficients are about 99\% along these degeneracy lanes. The locations of the ($a,b$) pairs, where the heating rate is roughly the same at both frequencies, follow the line $b=2.4-0.8a$, very similar to the degeneracy lane at 17\,GHz. Thus, in the case of the steady state EBTEL lookup table different metrics favor the best solution very similar to that for the \texttt{‘ebtel\_scale=0.2\_alpha=-2.5.sav’} lookup table. 

In contrast, the impulsive models, in particular the legacy impulsive lookup table \texttt{`ebtel.sav'} \citep{Nita2023}, do not provide a consistent solution, as illustrated in the bottom row of Figure\,\ref{Fig:metrics}; thus, they can be ruled out {for the mean coronal plasma component in the given AR instance. Note that, although the thermal model was fine-tuned using a limited number of voxels (those located along the isogauss surfaces corresponding to only two frequencies), the remaining coronal voxels associated with closed field lines are also filled according to the same parametric heating plasma model yielding a full 3D magneto-thermal model of this AR, as Fig.\,\ref{Fig:GR_layer_lines} illustrates}.

\section{Validating the thermal model with AIA data}\label{AIAvalidation}

We now check how closely our microwave-validated best model (that has been found in previous Section and is indicated in top row of Figure\,\ref{Fig:metrics} by red asterisks) reproduces emission in various coronal AIA channels. {We remind that the \mw\ emission is produced by thermal electrons, while the EUV emission is produced by ions, primarily highly ionized iron,  thus, for a given DEM, the AIA emission additionally depends on the elemental abundances \citep{2000ApJ...534L.203W}, which are known to vary in space and time \citep{2014ApJ...786L...2W,2015ApJ...802L...2C,2017ApJ...844...52D}. This implies an additional unknown in the model, which we do not investigate in this study by adopting standard coronal values \citep{1992PhyS...46..202F}.}
By construction, our best model describes only the averaged properties of the diffuse coronal plasma, as it heats the plasma at closed field lines only, according to  simplified formula (\ref{HeatingRate}). Thus, it does not include `selectively heated’ bright or dark EUV coronal loops and open field regions. In addition, the transition region (TR) contribution (so-called moss component) remains unconstrained because the microwave GR emission is almost insensitive to the TR plasma due to its low GR optical thickness. {Nevertheless, given that EBTEL accounts for the coupling between the corona and the lower atmosphere, the TR contribution from the involved TR pixels is defined provided that the EBTEL lookup table along with $Q_0$, $a$, and $b$ are known. The only remaining unknown is the set of the TR pixels that contribute to the moss component.  }

\begin{figure*}[!th]
\includegraphics[width=0.24\textwidth]{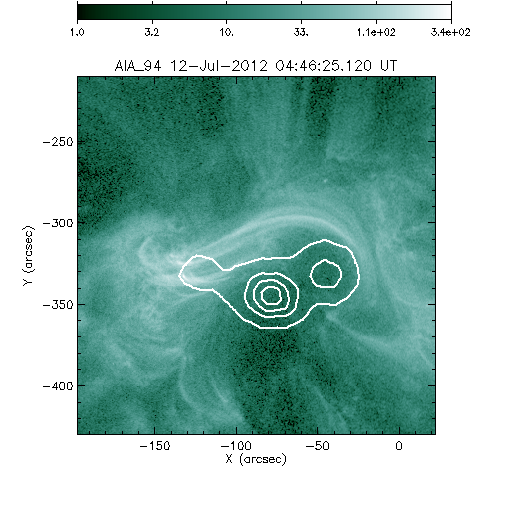}
\includegraphics[width=0.24\textwidth]{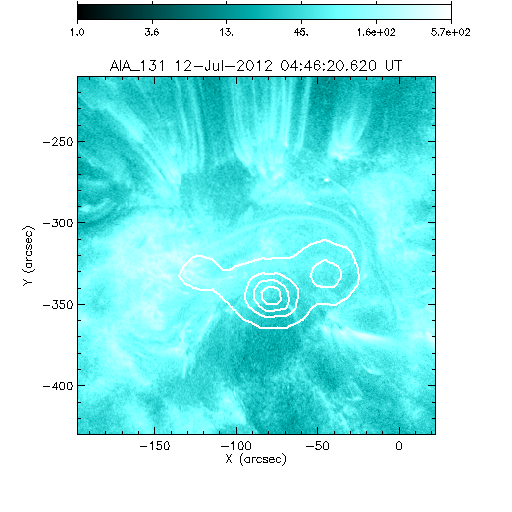}
\includegraphics[width=0.24\textwidth]{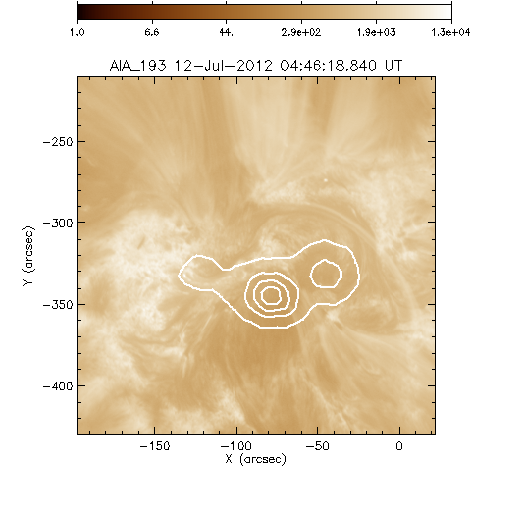}
\includegraphics[width=0.24\textwidth]{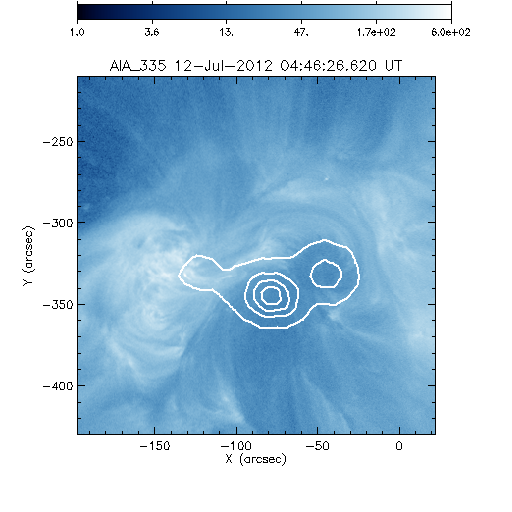}
\includegraphics[width=0.24\textwidth]{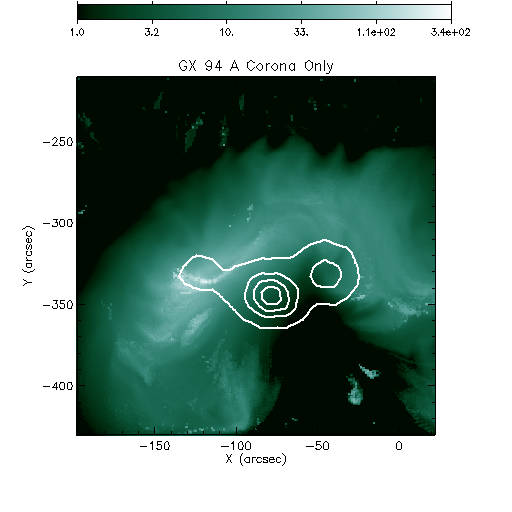}
\includegraphics[width=0.24\textwidth]{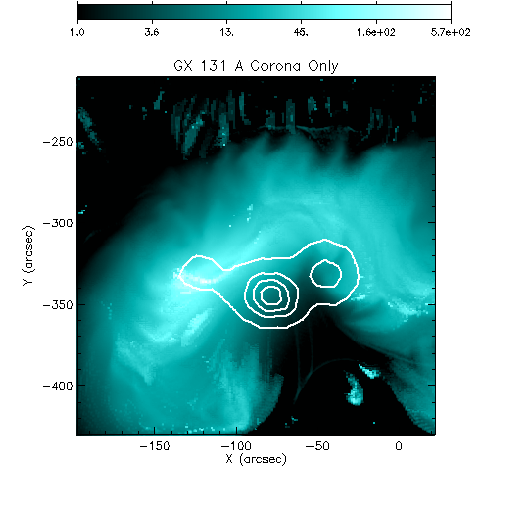}
\includegraphics[width=0.24\textwidth]{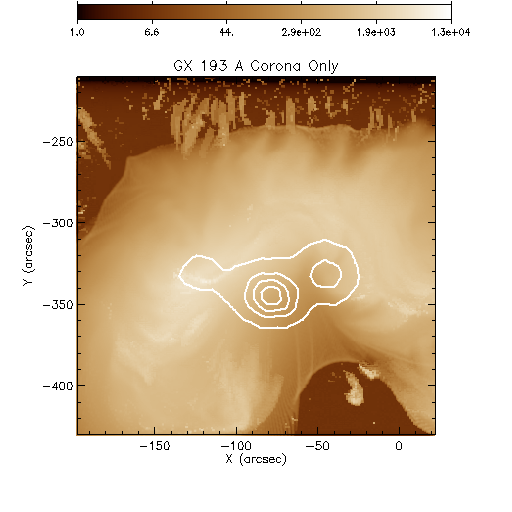}
\includegraphics[width=0.24\textwidth]{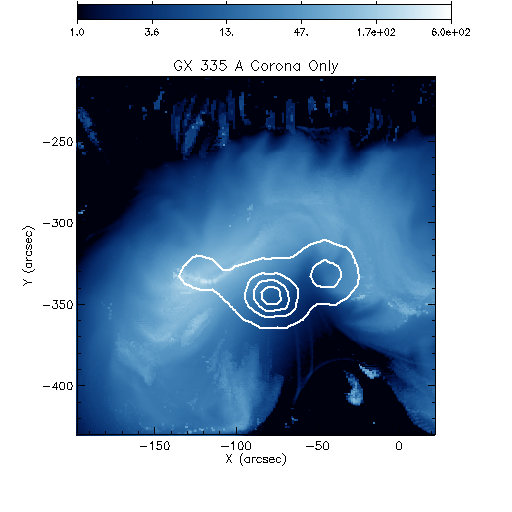}
\includegraphics[width=0.24\textwidth]{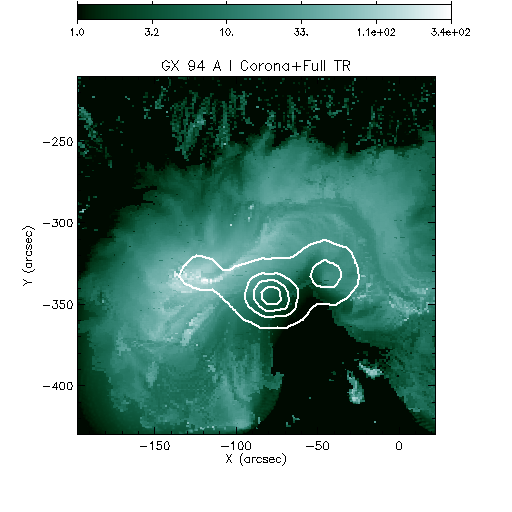}
\includegraphics[width=0.24\textwidth]{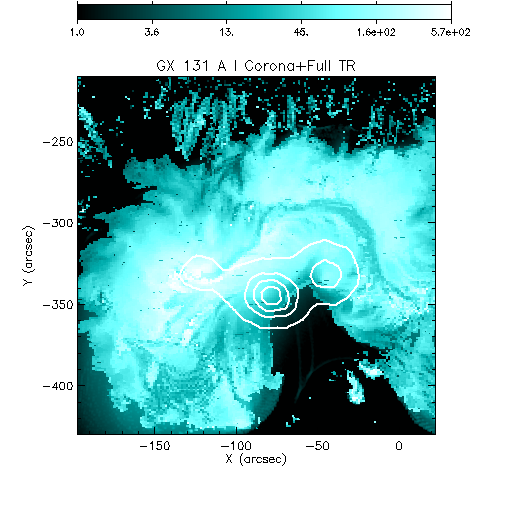}
\includegraphics[width=0.24\textwidth]{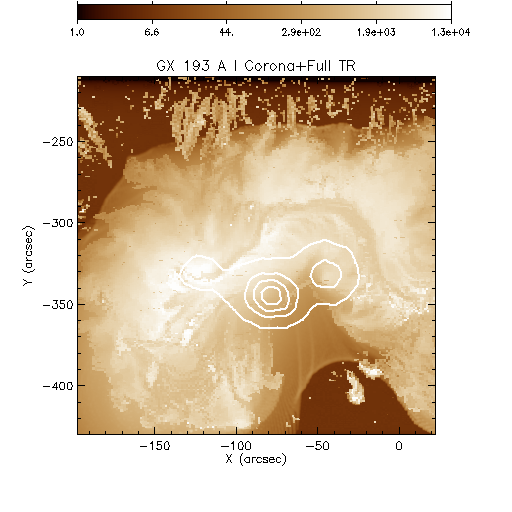}
\includegraphics[width=0.24\textwidth]{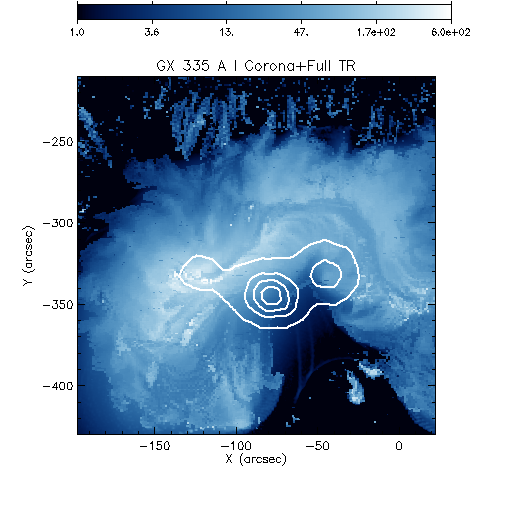}
\includegraphics[width=0.24\textwidth]{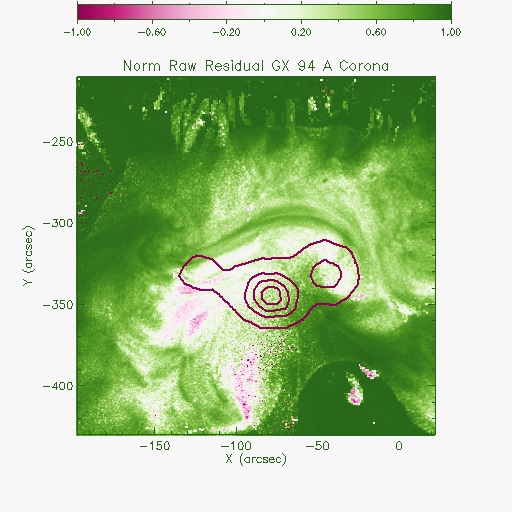}
\includegraphics[width=0.24\textwidth]{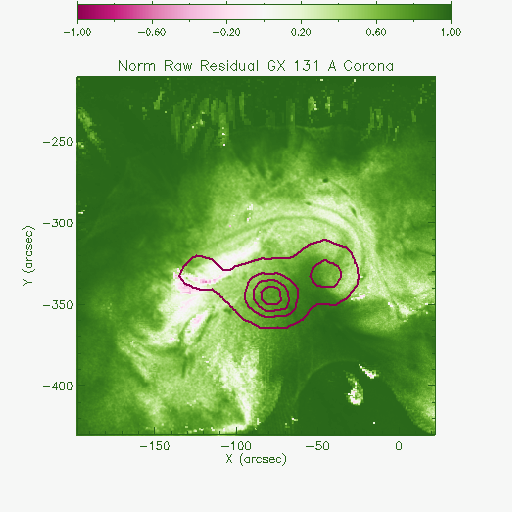}
\includegraphics[width=0.24\textwidth]{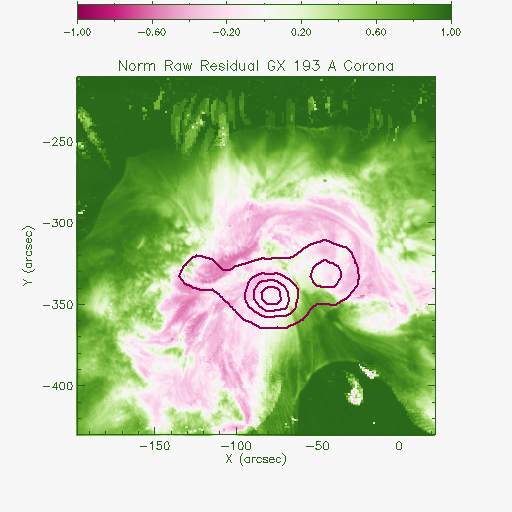}
\includegraphics[width=0.24\textwidth]{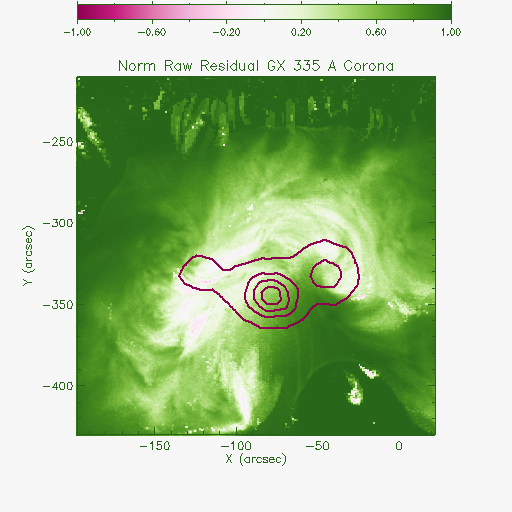}
\includegraphics[width=0.24\textwidth]{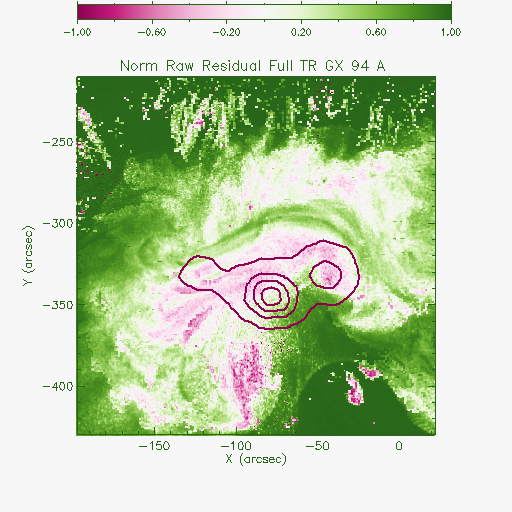}
\includegraphics[width=0.24\textwidth]{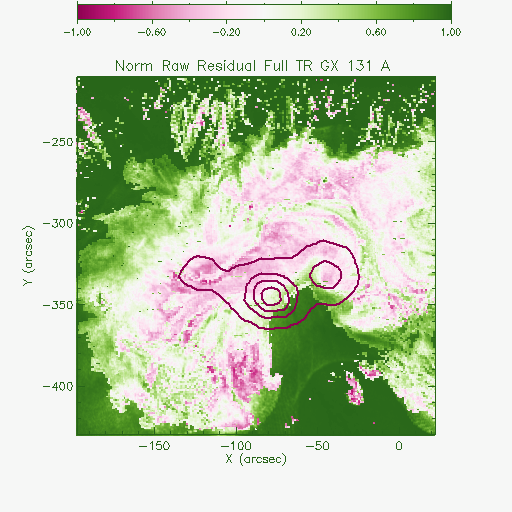}
\includegraphics[width=0.24\textwidth]{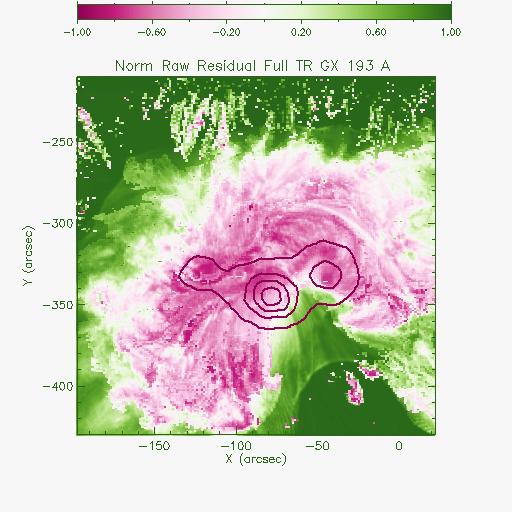}
\includegraphics[width=0.24\textwidth]{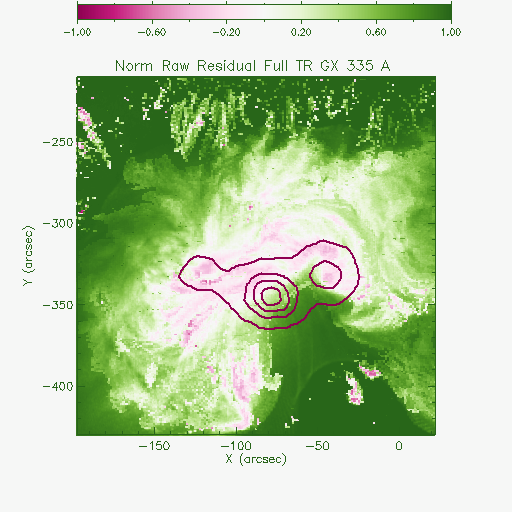}
\caption{Comparison of synthetic EUV emission computed using \mw-validated thermal coronal model in four AIA passbands with the observed AIA emission. The rows show: AIA image; synthetic coronal contribution; synthetic coronal and TR contribution; NORMALIZED residual between the AIA image and synthetic coronal image; and the same for synthetic coronal plus TR image. Contours of the \mw\ source at 17\,GHz are shown as a reference to outline the sunspots.}
\label{Fig:AIA_GX_all}
\end{figure*}

\begin{figure*}[!th]
\centerline{%
\includegraphics[width=\textwidth]{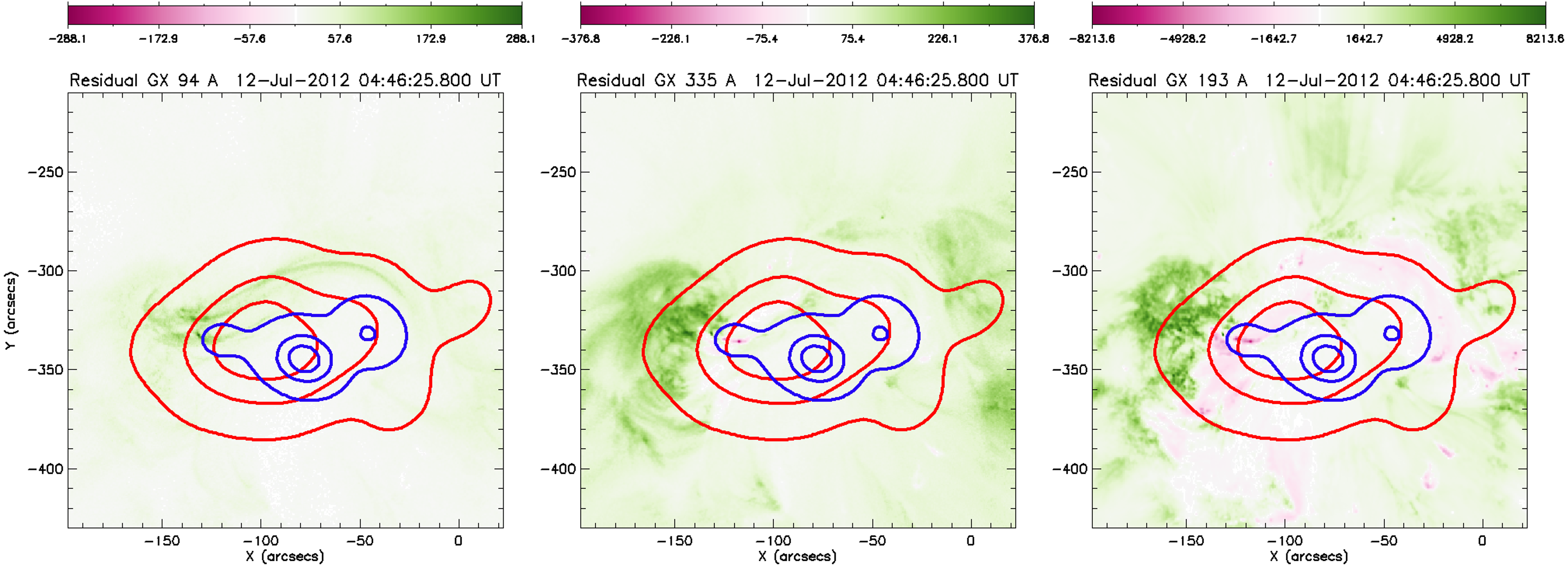}}%
\caption{EUV residual maps obtained by subtracting synthetic coronal images computed from the best thermal model fine-tuned with microwave data from the observed \SDO/AIA maps for AR 11520 for three representative passbands. A divergent color table was used to better indicate positive and negative residuals and areas with small residual. These plots indicate remarkably small residuals especially in the areas projected onto the microwave sources at 5.7 GHz (red contours) and 17 GHz (blue contours). In addition, they reveal the selectively heated bright or dark  loops, open loops, and TR patches. }
\label{Fig:AIA_res}
\end{figure*}

To assess our model versus AIA data, we synthesize the coronal AIA emission without TR contribution {and with the TR contribution from all TR pixels associated with closed magnetic field lines. For that, we} use our microwave-validated best model of the diffuse coronal component assuming standard coronal elemental abundances; {see Fig.\,\ref{Fig:AIA_GX_all}, second and third rows, respectively. Then, to compare the synthetic and observed data we form the normalized residuals between the AIA and synthetic images in the form of (data-model)/(data+model) and plot them in the fourth and fifth rows of Fig.\,\ref{Fig:AIA_GX_all}. These normalized residual plots demonstrate that the model strongly underestimates (dark green color) the AIA emission at the AR periphery. This is because the magnetic model lacks closed magnetic field lines at the data cube periphery, while the heating of the open field region is treated highly artificially in the GX Simulator. We focus on the AR core dominated by the closed field lines. There, the residual maps are dominated by white and light green colors indicating that
the coronal-only contribution is mostly insufficient to reproduce the observed AIA maps other than 193\,\AA, where the model slightly overestimates the data in the AR core (light red color), while a combination of the coronal and TR contributions overestimates the AIA emission in many locations; this effect is more pronounced in the cooler passbands. The reason for the excessive TR contribution is well understood and attributed to inability of the NLFFF models to correctly reproduce the magnetic field expansion that forms a ``canopy'' in the non-force-free low chromosphere \citep{Nita2018}. The TR ``moss'' component is observed to originate from strong-field regions; thus, to account for this effect,  \citet{Nita2018} introduced a TR mask, which permits the TR contribution only from a user-defined, presumably strong-field, TR locations. GX Simulator offers several approaches for this mask creation.

Normalized residual maps are highly sensitive to the locations where the emission is weak, while absolute residuals might be more representative in demonstrating the model success in reproducing AR bright core emission. }
A subset of the corresponding residual maps is presented in Figure\,\ref{Fig:AIA_res} for three AIA passbands. These maps demonstrate that the residuals are truly small in the AR core (white or light red/green colors indicating a mismatch within $\pm15\%$), where the model has been fine-tuned using the microwave data, which indicates a remarkable success of this modeling approach. {Notably, the area of small residuals, i.e., of good model-to-observations agreement, extends far beyond the sunspots (indicated by the 17 GHz intensity contours), which means that the microwave observations allow us to constrain the thermal plasma parameters not only in their sources, but within (almost) the entire AR as well.} However, the residual maps also reveal other mentioned components---TR patches (``moss'', at the left and right edges of the second and third panels), selectively heated loops (bright (green) at 94 {\AA} {also seen in the normalized residual maps shown in Fig\,\ref{Fig:AIA_GX_all}}, while dark (red) at others), and open field regions (in the upper-right corner of the third panel). It is straightforward to separate those components in the residual maps and then analyze them individually. 

\begin{figure*}[!th]
\centerline{%
\includegraphics[width=0.99\textwidth]{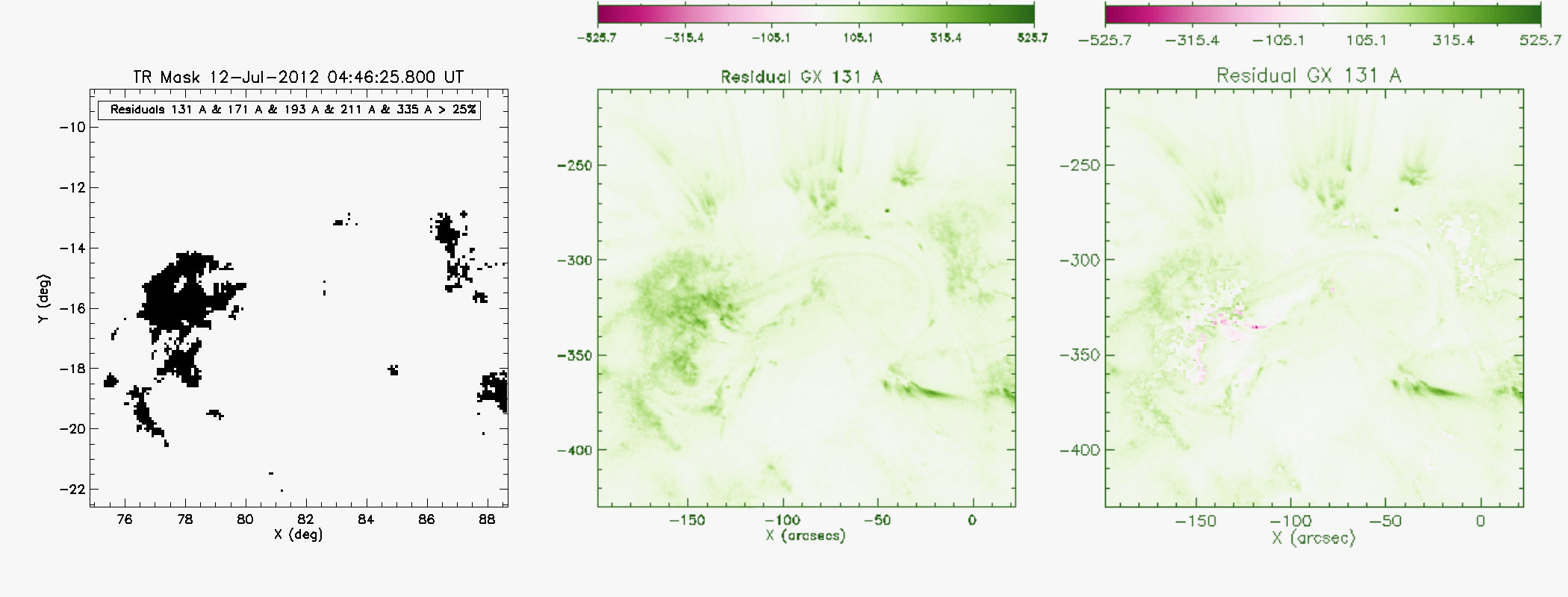}}%
\caption{Left panel shows a TR mask computed as a collection of pixels where the residuals between the AIA images and synthetic coronal images exceed 25\% of the image peak values simultaneously at 131, 171, 193, 211, and 335 \AA.  Middle: 131 {\AA} residual map (observations minus model); no TR contribution has been added to the synthetic image. Right:  131 {\AA} residual map; a TR contribution has been added to the synthetic image according to the mask shown in the left panel. Significant improvement (reduction) of the residual is apparent.}
\label{Fig:AIA_res_TRmask}
\end{figure*}


The TR component is seen in the form of similarly located patches in several residual maps, most notably 131, 171, 193, 211, and 335 {\AA}, while the coronal loop structures have been found to have noticeably different shapes at different passbands. This implies that the TR component can be isolated by making a cumulative mask covering several mentioned residual maps. To this end, GX Simulator now permits one to make a TR mask by marking all image pixels, where the residual exceeds a user-defined threshold, for example 25\% of the peak values, at several user-selected residual maps. This way, only the pixels showing the excess at several residual maps are added to the mask. 
Figure\,\ref{Fig:AIA_res_TRmask} shows an example of such a 25\% mask. 
{Although this mask was selected “by eye” rather than through algorithmic optimization, this rough choice nonetheless clearly demonstrates that including the TR contribution from the masked pixels---computed using the same heating model and EBTEL lookup table as for the diffuse coronal component---leads to a measurable improvement in the model-to-data agreement, as evidenced by a significant reduction in the residuals compared to both the coronal-only and the full TR-inclusive models shown in Fig.~\ref{Fig:AIA_GX_all}.} Knowing the areas producing dominant TR contribution can help study those areas in their association with underlying photospheric/chromospheric conditions, including the magnetic field, brightness, and velocities. {The remaining residuals (that are unaccounted for by even the masked TR model) can be attributed to local deviations from the ``global'' coronal heating model (\ref{HeatingRate}), e.g., by selectively heated loops; such deviations, however, seem to be isolated and relatively weak.}

\section{Discussion}

Our analysis vividly demonstrates quantitative success of the parametric heating modeling in reproducing the thermal structure of an AR. Indeed, it is capable of faithfully reproducing the diffuse (mean) component of the coronal AR plasma and of identifying the TR moss component of the AR. Somewhat unexpected is the finding that a high-frequency nanoflare model or a steady-state one offer a much better approximation to the reality than the lower-frequency (impulsive) nanoflare models. This is in apparent contradiction with the finding of \citet{2017ApJ...842..108V} made using a time lag technique favoring the low- or moderate- frequency nanoflare heating. \citet{2024arXiv241220348M} proposed that the nanoflare frequency is a function of location within the AR and of its evolution stage, so all low-, moderate-, and high- frequency nanoflares may play a role. Yet, they found that the AR core heating is dominated by a low-frequency heating, while in our case the heating model fine-tuned in the AR core favors the high-frequency / steady-state heating. It must be noted, however, that the direct comparison can be biased: EUV analysis was typically performed on isolated bright (selectively heated) loops, while our conclusions are relevant for the diffuse coronal component, whose heating regimes might be different from each other. In principle, the coronal plasma properties might depend on free energy accumulation and flaring release. In our case, AR 11520 has likely accumulated a lot of free energy, which was released several hours after our data in the form of an X-class flare \citep{Liu_2022}.

Our residual map inspection allowed us to isolate several distinct components of the coronal thermal structure of the AR: the TR moss component, selectively heated loops, and open field loops. The bright TR areas (moss component) can straightforwardly be accounted for within the model by creating the corresponding TR mask and adding TR contribution described by the same EBTEL lookup table, with the same model parameters $Q_0$, $a$, and $b$ as inferred from the \mw\ data; see Figure\,\ref{Fig:AIA_res_TRmask}. Two other identified components, the selectively heated and open loops, are coronal and cannot be recovered within the adopted here simplified approach to the coronal heating model described by Eqn. (\ref{HeatingRate}). The selectively heated loops, such as those apparent from Figure\,\ref{Fig:AIA_res}, left panel, require some additional heating compared to that responsible for the diffuse component, provided that the required connectivity is available in the magnetic model. If it is not available, than the magnetic model itself needs corrections to recover this connectivity and improve the final magneto-thermal model. In principle, such an improvement can be attempted by using time-dependent magneto-frictional  or data-driven  models \citep[see, e.g.,][]{Liu_2022,2023ApJ...952..136A,Fan_2024}. 

How and why the bright coronal loops are heated differently from the surroundings is a crucial yet unanswered question in solar physics. 
One promising option permitting a selective heating of coronal loops is the dependence of the heating on the photospheric connectivity of the coronal loops discovered by \citet{2017ApJ...843L..20T}. They  demonstrated that the loops with both footpoints rooted in the dark umbra are invisible in EUV. In contrast, if one footpoint is rooted in umbra, while the other one is in penumbra or plage, such loops are the brightest. The loops with both footpoints rooted in plages are brighter than the background, but they are never the brightest ones. All these findings are in line with the idea that the magnetoconvection at the photosphere, which presumably drives the coronal magnetic line braiding, plays a key role in the coronal heating. To account for the connectivity effect, the model heating has to permit additional (\textit{a priori} unknown) numeric factors to the volumetric heating rate that enhance or reduce the heating compared to an ``average'' one, depending on the locations of the involved footpoints, a functionality not yet available in our search engine. In addition, the selectively heated loops can host a different heating regime (simplistically speaking, a different EBTEL lookup table) compared with that in the diffuse component. {It is is important to realize that these selectively heated loops can be brighter than the diffuse component in some AIA passbands, while darker in others. Perhaps, having the properly tuned mechanism of the selective heating can solve the problem of the slight overestimation of the 193\,\AA\ emission identified in Fig.\,\ref{Fig:AIA_GX_all}, fourth row. Alternatively, this might require an adjustment (reduction) of the adopted elemental abundances.
}

The identified open-field regions may contain both closed loops, whose second footpoint is outside the modeling box and truly open field lines. For the former case, we can, in principle, add a heating with a minor modification to our parametric heating model by guessing (extrapolating) what is the field line behavior outside the box. For the latter, the heating mechanism itself can be highly different from that of the closed field lines \citep[see, e.g.,][]{2021SoPh..296..144V}, which can call for a different treatment of the open field line heating \citep[e.g.,][]{2014ApJ...787..160W}.

To conclude, the parametric coronal heating approach that employs an NLFFF magnetic model of an AR dressed using solutions of the hydrodynamical code EBTEL, combined with both \mw\ multi-frequency maps (not sensitive to elemental abundances) and EUV imaging data (sensitive to the elemental abundances), permits obtaining a truthful magneto-thermal model that closely matches the available observational constraints. {Although a slight reduction in elemental abundances might be necessary to reproduce the 193\,\AA\ emission, this conclusion remains premature unless the selective heating has been properly incorporated into our model. Thus, no significant deviation from the standard coronal abundances can currently be reported for the AR 11520 considered in this study.}\\

This work was supported in part by NSF grant 
AST-2206424,  
and NASA grant 
80NSSC23K0090 
to New Jersey Institute of Technology (GDF and GMN), and by the RF Ministry of Science and Higher Education (AAK). 

\bibliographystyle{aasjournal}
\bibliography{CoronalHeating,cdis}

\end{document}